\documentclass[runningheads]{llncs}
\usepackage{graphicx}

\usepackage{multirow,url}
\usepackage{tikz,msc,color,pifont}
\usetikzlibrary{arrows.meta}
\usetikzlibrary{fit,backgrounds}
\newif\ifediting
\editingtrue  

\newcommand*\circled[1]{%
  \tikz[baseline=(C.base)]\node[draw,circle,inner sep=1.2pt,line width=0.2mm,](C) {#1};}
\newcommand*\Myitem{%
  \stepcounter{enumi}\item[\circled{\theenumi}]}

\begin{document}
\title{Can Authoritative Governments Abuse the Right to Access?}
%
\author{C\'edric Lauradoux\inst{1}}
\authorrunning{C. Lauradoux}
%
\institute{University Grenoble-Alps, INRIA, Grenoble, France \email{cedric.lauradoux@inria.fr} }
\maketitle              
\begin{abstract}
The right to access is a great tool provided by the GDPR to empower data subjects with their data. However, it needs to be implemented properly otherwise it could turn subject access requests against the subjects privacy. Indeed, recent works have shown that it is possible to abuse the right to access using impersonation attacks. We propose to extend those impersonation attacks by considering that the adversary has an access to governmental resources. In this case, the adversary can forge official documents or exploit copy of them. Our attack affects more people than one may expect. To defeat the attacks from this kind of adversary, several solutions are available like multi-factors or proof of aliveness. Our attacks highlight the need for strong procedures to authenticate subject access requests.  

\end{abstract}
\section{Introduction}

In 2011, Max Schrems, an Austrian citizen, sent a request to the US company Facebook\footnote{\url{http://europe-v-facebook.org}} to access all the data held by the company. He received a huge PDF file (1200 pages) describing the data Facebook held on him. The analysis of the data shows that Facebook was not respecting European data protection laws. In 2018, David Caroll submitted a subject access request to UK company Cambridge Analytica. The company first refused to comply with Caroll's request, and it escalated until the UK data protection authority~\cite{ICO2018} asked the company to comply with Caroll's access request. In the end, Cambridge Analytica's actions  during 2016's US election were exposed. In both cases, Facebook and Cambridge Analytica had to provide data to individuals in order to respect their right to access. In Europe, this right is defined in Article 15 of the GDPR.  The right to access data is considered as the root for all the other rights defined by the GDPR. Once, a data subject has accessed the data collected by a data controller, he/she can exercise the other rights (to object, to data portability, to be forgotten, etc.). 

The right to access needs to be implemented carefully to avoid data breaches. Indeed, an adversary can send a subject access request while impersonating a legitimate subject to a data controller. If the data controller does not authenticate the access request properly, the adversary can obtain the data by abusing the right to access. This threat has to be considered seriously as data protections officers (DPOs) have already reported in~\cite{Martino2022} suspicious 
access requests. Researchers~\cite{Boniface2019,Bufalieri2020,Cagnazzo2019,Martino2022,Martino2019,Pavur2019} have conducted studies to test if the right to access was implemented properly. They have used different strategies~\cite{Bufalieri2020,Cagnazzo2019,Martino2019,Pavur2019} to fool DPOs and succeeded to steal personal data in numerous cases. In response to those issues, the European Data Protection Board (EDPB) has recently published some guidelines~\cite{EPDB2022} on the right of access to clarify its implementation and to improve the situation. Still many things have to be done to fix how subject access requests are handled by data controllers. 

In this work, we focus on data controllers who authenticate a subject access request using an official document like an ID card or a passport. This authentication is often encountered in practice despite the fact that it can be irrelevant as pointed out in~\cite{Boniface2019,EPDB2022}. We alert that there are more elaborated attacks possible if the adversary benefits from the support of a government. In this case, the adversary can forge official document and impersonate any data subjects to forge subject access request. It allows authoritative or corrupted states to abuse the right to access to spy on dissidents who use online services eligible to subject access request. This kind of attacks needs to be avoided because it could clearly harm the reputation of the data protection regulations like the GDPR. 

Official document forgery shows the limit of document-based authentication. Additional measures needs to be implemented by data controllers to prevent ID forgery. Multiple-factor  authentication (MFA) is an interesting solution to thwart forgery attacks. It requires that the data subject and the data controller agree on an additional secure communication (an electronic mail address or a phone number).

The paper is organized as follows. Section~\ref{sec:implementation} discusses how data controllers authenticate subject access requests. We describe the benefit of implementing the right to access using privacy dashboard or by DPOs. We review the existing attacks against the right to access in Section~\ref{sec:threats}. We introduce our new attack scenarios in Section~\ref{sec:new}. Solutions are discussed in Section~\ref{sec:counter}.

\section{How are subject access requests authenticated?}\label{sec:implementation}

The right to access one's personal data (Article 15 of GDPR~\cite{EU-679-2016}) via Subject Access Requests (SAR) is fundamental to ensure data protection. It makes the processing of personal data by organizations transparent and accountable towards data subjects, whereby these are made aware of, and can verify the lawfulness of the processing of their data (Recital 63 of GDPR~\cite{EU-679-2016}).

This right to access by individuals complements the mandate of the data protection authorities to monitor and enforce the application of the GDPR (Article 57.1.a of GDPR~\cite{EU-679-2016}). The right to access also enables the exercise of all the other rights (erasure, rectification, restriction, etc). When a data controller receives an access request, it needs to perform two types of verification. A  verification of the \emph{eligibility} needs to be done. A data controller needs to verify if the subject sending a request is concerned by a data protection regulation which includes a right to access. A verification of the \emph{legitimacy} needs also to be done by the data controller. Verifying the legitimacy has two purposes: i) establish the subject's identity, and ii) check that the subject's identity matches the requested data. Verifying both \emph{eligibility} and \emph{legitimacy} can be obtained by executing an appropriate authentication protocol between the data subject sending the access request and the data controller. 

If no verification is implemented, the data of a subject can be collected by anyone impersonating the subject to the data controller. This is particularly demonstrated by Bufalieri \textit{et al.}~\cite{Bufalieri2020}. They submitted 334 subject access requests to different data controllers, and 58 organizations provided data without any verification of the subject sending the request. It confirms that the right to access without authentication  is harmful to privacy. We discuss next how data controllers are actually handling subject access requests.

\subsection{Dashboard versus DPO}

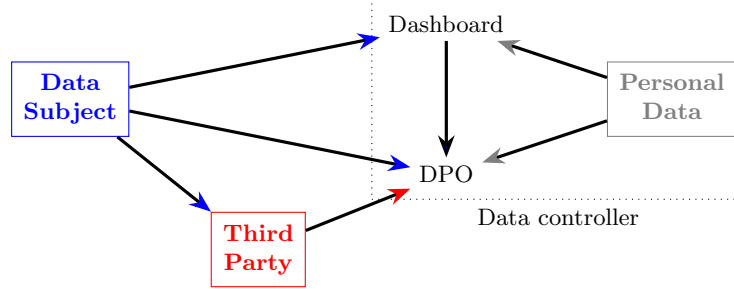
\begin{figure}[!ht]
\centering
\caption{Current options for the implementation of the right to access.}\label{fig:methods}
\begin{tikzpicture}
\begin{scope}[]node/.style={rectangle,thick,draw,rounded corner}]
   
    \node[draw=blue,rectangle] (A) at (0,0) {\textcolor{blue}{\textbf{\begin{tabular}{c}
    Data\\Subject
    \end{tabular}}}};
    \node(B) at (5,1) {Dashboard};
    \node (C) at (5,-1) {DPO};
    \node[draw=gray,rectangle] (D) at (8,0) {\textcolor{gray}{\textbf{\begin{tabular}{c}
    Personal\\Data
    \end{tabular}}}};
	\node[draw=red,rectangle] (E) at (2.5,-2) {\textcolor{red}{\textbf{\begin{tabular}{c}
    Third\\Party
    \end{tabular}}}};
     \node[draw,dotted,fit=(B) (C) (D),label=below:Data controller] {};
\end{scope}
    
\begin{scope}[
              every node/.style={fill=white,circle},
              every edge/.style={draw,very thick}]
    \path[draw=black,fill=blue] [-{Stealth[slant=0]}] (B) edge (C); 
    \path[draw=blue,fill=blue] [-{Stealth[slant=0]}] (A) edge (B); 
    \path[draw=blue,fill=blue] [-{Stealth[slant=0]}] (A) edge(C); 
	\path[draw=blue,fill=blue] [-{Stealth[slant=0]}] (A) edge(E);   
	\path[draw=red,fill=red] [-{Stealth[slant=0]}] (E) edge (C); 
	\path[draw=gray,fill=gray] [-{Stealth[slant=0]}] (D) edge (B); 
	\path[draw=gray,fill=gray] [-{Stealth[slant=0]}] (D) edge (C); 
\end{scope}
\end{tikzpicture}
\end{figure}

Two methods to handle subject access requests (see Fig.~\ref{fig:methods}) have been implemented by data controllers. In the first method, a privacy dashboard has been implemented by the data controller. The subject authenticates to the controller's website and then can obtain his/her data or submit securely an access request to the controller's DPO. This solution has been chosen by Google or Facebook. They have created dashboard on their websites to let data subjects access their data. To access the dashboard, the data subject needs first to authenticate to the data controller websites. An extra verification may be required to obtain the data from the data controller as part as a two-factor authentication. Privacy dashboards  are very convenient for the data subjects because the creation and the processing of their request is automatic. Google Takeout (\url{https://takeout.google.com}) is a good example of dashboard which implements two-factor authentication. More details on privacy dashboards can be found in~\cite{Tolsdorf2021}.  

In the second method, all the subject access requests are handled by the controller's DPO. The DPO is in charge of verifying both eligibility and legitimacy of the subject access requests. In this method, a first issue is to setup a secure communication channel to let the subject submits his/her request to the DPO. It can be done through a web page or by sending 
and electronic mail\footnote{Other more classical methods like postal mail are also available but they are not considered in this work.}. In any case, the communication channel between the subject and the DPO needs to be secure (using end-to-end encryption for instance). Once a secure channel has been setup, the DPO can interact with the subject to verify his/her request.  It is important to notice that third parties can help the subject to contact the DPOs. A data protection authority can be contacted in the case for an indirect access. It can be also a third party which helps subjects to create their subject access requests like Tapmydata\footnote{\url{https://tapmydata.com}}.

In Table~\ref{tab:comparison}, we compare the main characteristics of the different implementations one can come across. The security of the authentication scheme used in a dashboard is based on a secret shared (\textit{i.e.} a password) between the data controller and the data subject. It is human-to-computer security mechanism based on the following assumptions: (i) the data subject does not share the secret with other individuals, (ii) the secret is not re-used to other data controllers, (iii) the secret is difficult to guess and (iv) the authentication is performed over secure channel. This technology is tried-and-tested even if it has some limitations~\cite{Bonneau2015}.

\begin{table}[!ht]
\centering
\caption{Comparison of the properties of the different implementations of the right to access.}\label{tab:comparison}
\begin{tabular}{|r|r|r|r|}
\hline
                           & \textbf{Dashboard}          & \textbf{DPO}         \\  \hline
\textbf{Creation of SAR}   &     Automatic               & Manual            \\ 
\textbf{Processing of SAR} &     Automatic/Manual               & Manual            \\
\textbf{Security model}    &     Password (+MFA)      & Choice of the DPO            \\
 \hline
\end{tabular}
\end{table}

The authentication scheme implemented by DPO is a human-to-human interaction. It is based on the assumptions that: (i) the communication channels used between the subject and the DPO are secure, (ii) the data subject is the only individual able to prove that he/she has created his/her request and (iii) the DPO is trained to verify requests.  The verification is a choice made by the DPO. This choice is critical to authenticate subject access requests. DPOs often ask the subject sending a request to provide a copy of an official identification document. This method of verification has been criticized in~\cite{EPDB2022}.

\section{Threat Model and Known Attacks}\label{sec:threats}

A general threat model for the right to access was proposed by Boniface \textit{et al.}~\cite{Boniface2019}. This model considers both malicious subjects and malicious controllers, whereby three different categories of threats were defined: i) denial of access, ii) privacy invasion and iii) data breach due to impersonation. We briefly review denial of access and privacy invasion and then focus on data breach.

\paragraph{Denial of access.} This situation corresponds to cases in which data controllers decline to provide data to data subjects; one alleged reason refers to scenarios wherein they are not in a position to verify the identity of subjects.It also occurs often with companies collecting and processing cookies~\cite{Boniface2019,Degeling2019}, but also with those processing IP addresses~\cite{Adhatarao2021}. Denial of access was also observed in~\cite{Ausloos2018,Boniface2019,Degeling2019,Urban2019}. 

\paragraph{Privacy invasion.} This attack is defined in~\cite{Boniface2019} as a case in which the data controller is honest-but-curious but requests additional information to authenticate the SAR. These additional requested information can increase the exposure of the subject to the controller and potentially violates the \emph{minimization and proportionality principles}  prescribing that personal data shall be is adequate, relevant and limited to what is necessary in relation to the purposes for which they are processed (Article 5 (1)(c) of the GDPR).
For example, the controller should not be obliged to acquire additional information in order to identify the data subject for the sole purpose of being complaint (Recital 57 of the GDPR).

\paragraph{Data breach.} Data breach refers to the case of a data controller which provides the data of a given subject A to another subject B. A data breach can be the result of an error or it can be the result of an impersonation attack. Example of impersonation attacks against the right to access are given in~\cite{Bufalieri2020,Cagnazzo2019,Martino2019,Pavur2019}. These studies have demonstrated that it is possible to use the right of access to steal data by impersonating a data subject. Recent studies~\cite{Bufalieri2020,Cagnazzo2019,Martino2019,Pavur2019} show that data controllers are not ready to process correctly subject access requests because they still use weak authentication procedures vulnerable to impersonation attacks. Table~\ref{fig/attack} summarizes the different methods used to impersonate a subject and to abuse  data controllers with illegitimate subject access requests. It is important to notice that the attacks were successful against data controllers which rely their DPOs to authenticate subject access requests. 

\begin{table}[!ht]
\centering
\begin{tabular}{|r|r|r|}
\hline
\textbf{Authentication method} & \textbf{Attacks} & \textbf{Target} \\ \hline
\multirow{2}*{Copy of an ID}        & Social engineering~\cite{Pavur2019} & DPO\\
         & Falsification~\cite{Martino2019} & DPO \\ \hline

Email confirmation    & Email  spoofing~\cite{Bufalieri2020,Cagnazzo2019,Martino2019} & DPO\\ \hline
Bills    & Falsification~\cite{Pavur2019} &DPO \\  \hline
Personal question   & Social engineering~\cite{Pavur2019} & DPO\\
\hline

\end{tabular}
\caption{Known impersonation attacks abusing the right to access and their implementation.}\label{fig/attack}
\end{table}

\section{Advanced Forgery Attacks}\label{sec:new}

\begin{figure}[!ht]
\centering
\begin{tikzpicture}[scale=0.78]
\begin{scope}[every node/.style={circle,thick,draw}]
   
    \node [color=red](B) at (-2,3) {\textbf{Charlie}};
    \node [color=red](C) at (4.8,4.5) {\textbf{Eve}};
    \node (D) at (4.8,1.5) {\textbf{Alice}};
    \node (E) at (10,3) {\textbf{Bob}};

\end{scope}
\begin{scope}[>={Stealth[red]},
              every node/.style={fill=white,circle},
              every edge/.style={draw,very thick}]

\path [->,color=red,arrows={->[red]}] (B) edge[bend left=100] node {\begin{small}\textcolor{red}{\ding{172}
 Identification}\end{small}} (E);

     \path [->,color=red] (C) edge[bend left=30] node {\begin{small}\begin{tabular}{c}\ding{175} Alice's\\data\end{tabular}\end{small}} (B);

       \path [->,color=red,arrows={->[red]}] (B) edge[bend left=30] node {\begin{small}\textcolor{red}{\begin{tabular}{c} \ding{173} Forged\\Alice's ID\end{tabular}}\end{small}} (C); 
    
    \path [->,color=black,arrows={->[black]}] (B) edge[bend right=30] node {\begin{small}\begin{tabular}{c} Alice's\\ID\end{tabular}\end{small}} (D); 
    
    \path [color=red,arrows={->[red]}] (C) edge[bend left=30] node {\begin{small}\begin{tabular}{c}\ding{174} SAR using\\Alice's ID\end{tabular}\end{small}} (E); 
    
    \path [color=black,arrows={->[black]}] (E) edge[bend left=30] node {\begin{small}\begin{tabular}{c}\ding{175} Alice's\\data\end{tabular}\end{small}} (C);

\end{scope}
\end{tikzpicture}
\caption{Impersonation attack by an authoritative state (Charlie) against a dissident (Alice) using a proxy (Eve) to contact a data controller (Bob).\label{fig:authorative}}
\end{figure}
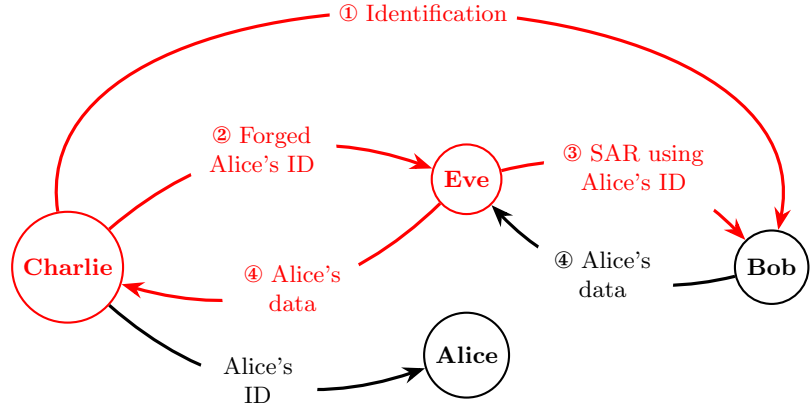

We demonstrate in this section that verification of subject access request based on official identification document can not resist to an adversary which is backup by a government. Indeed, we focus on DPOs who verify subject access requests by asking a copy of official identification document. Despite being criticized by EDPB in its guidelines~\cite{EPDB2022}, this method of verification is still considered valid depending on the information known by the data controller on the data subject. We assume that if the adversary is able to provide an official identification document then he/she can fool DPOs and therefore abuse the right to access. Other authors (like~\cite{Martino2022} for instance) have made the same conclusion but they have not described scenarios and their impact as it is done in our work. Our attack is foreseen in three steps (as depicted in Figure~\ref{fig:authorative}). 
\begin{enumerate}

\Myitem The malicious state (represented by Charlie) needs first to identify the websites and services used by the target (Alice)  located in the EU. He obtains the contact information of all the data controllers collecting and processing the data of their targets (including Bob). 

\Myitem Charlie produces official documents (ID card, passport\ldots) which match the identity of Alice. Charlie creates documents for a perfect namesake with the same first, last name, birth date, gender, etc. These documents are transmitted through a proxy (Eve). 

\Myitem Eve submits the subject access request to Bob and to any other controllers. Eve can provide valid government-issued documents whenever they are requested by  data controllers. 

\Myitem Eve (and later Charlie) obtains Alice's data from Bob.    
\end{enumerate}

The first step is not mandatory. Indeed, Charlie and Eve can adopt a guess-and-determine strategy and submit subject access requests to many data controllers without knowing whether they collect and process personal data from Alice. If the data controller (Bob) answers, Charlie and Eve obtain Alice's data. However, this strategy is not particularly discreet and some data controllers may alert Alice that they have received strange requests. 

The authoritative state may not necessarily need to use Eve as a proxy. Eve can be useful since she can be located in the same country as the targeted dissident: the request submitted by Eve would mimic the request that the target could have submitted. 

Tools like the application Tapmydata\footnote{\url{https://tapmydata.com/}} can simplify the task of the adversary. Tapmydata can be used to submit automatically requests to data controllers.  It aims at simplifying the creation and the submission of subject access requests. As such, an adversary can register with an email address and confirm her identity by uploading a scanned of his/her forged passport. The application knows how to contact 1400 data controllers and takes care of all the administrative details of the request. When a data controller replies to a request, the data is stored in a docker. In our context, it amplifies our attack. 

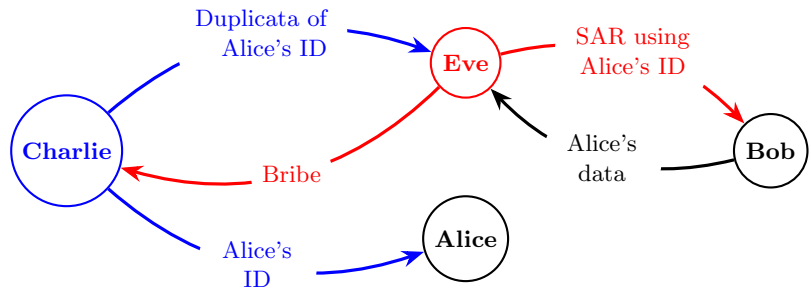
\begin{figure}[!ht]
\centering
\begin{tikzpicture}[scale=0.78]
\begin{scope}[every node/.style={circle,thick,draw}]
   
    \node [color=blue](B) at (-2,3) {\textbf{Charlie}};
    \node [color=red](C) at (4.8,4.5) {\textcolor{red}{\textbf{Eve}}};
    \node (D) at (4.8,1.5) {\textbf{Alice}};
    \node (E) at (10,3) {\textbf{Bob}};

\end{scope}
\begin{scope}[>={Stealth[red]},
              every node/.style={fill=white,circle},
              every edge/.style={draw,very thick}]
    
     \path [->,color=red] (C) edge[bend left=30] node {\begin{small}\textcolor{red}{Bribe}\end{small}} (B);

       \path [->,color=blue,arrows={->[blue]}] (B) edge[bend left=30] node {\begin{small}\textcolor{blue}{\begin{tabular}{r}Duplicata of\\Alice's ID\end{tabular}}\end{small}} (C); 
    
    \path [->,color=blue,arrows={->[blue]}] (B) edge[bend right=30] node {\begin{small}\begin{tabular}{c}Alice's\\ID\end{tabular}\end{small}} (D); 
    
    \path [color=red,arrows={->[red]}] (C) edge[bend left=30] node {\begin{small}\begin{tabular}{c}SAR using\\Alice's ID\end{tabular}\end{small}} (E); 
    
    \path [color=black,arrows={->[black]}] (E) edge[bend left=30] node {\begin{small}\begin{tabular}{c}Alice's\\data\end{tabular}\end{small}} (C);

\end{scope}
\end{tikzpicture}
\caption{Impersonation attack when an administrative Charlie is corrupted by Eve.\label{fig:corrupted}}
\end{figure}

A similar scenario can occur if an administrative who works in a department delivering official identification documents is corrupted. Figure~\ref{fig:corrupted} illustrates this case wherein Alice is a data subject using an online social network maintained by Bob (and located in the EU). Eve wants to obtain information on Alice. Eve knows that Charlie can be corrupted. Eve bribes Charlie to obtain the genuine identification document bearing Alice's information. Eve submits a SAR to Bob impersonating Alice. Eve can provide all the identification documents. 

\subsection{Impact Assessment}

One may consider that this attack is unlikely to occur and that the adversary model is too strong. Governments have other legal means to access data from their citizens. Our attack scenario makes sens if we consider the case of dissidents and authoritative states. Let us assume that some dissidents are using online services located in Europe. The authoritative state they are against can use our forgery attack to abuse the right to access and spy on them. This scenario would demonstrate that the GDPR can be abused for surveillance purpose. Creating on-demand official identification documents and sending subject access requests are rather inexpensive. If the impersonation succeeds, there are two types of consequences: 

\begin{itemize}
    \item Direct consequences: a state can destabilize, denigrate or blackmail dissidents online. 
    
    
\item Indirect consequences: 
after a first impersonation attack, a state can further impersonate the target to exercise other rights (depending on its goal), for example: 
 
\begin{itemize}
\item  rectification (Article 16) to modify data or content published by the target. It can harm the reputation of the target or it can be used to denigrate his/her speech.   

\item erasure (Article 17) as an effective censorship method by removing all the data of the dissident from a website (his/her blog for instance). This is a form of denial of service attack. 

\end{itemize}    
    
\end{itemize}

The scope of the GDPR increases the impact of our attack to more categories of people. First, we have dissidents who are using European online service. The authoritative state they are against can forge official identification documents at will and submit subject access request to data controllers which are in EU. Second, European citizen who have visited an authoritative state. The authoritative state can keep a copy of their passport when they cross the border during the customs check. The authoritative state can submit subject access requests to data controller anywhere who collect and process the data of the target. Third, Foreigners (Non European) who are using European online service who have visited an authoritative state. It demonstrates that an authoritative state can abuse the right to access and exploit the scope of the GDPR to reach as many people as possible.

\subsection{Ethical considerations}

We have not implemented our attack due to ethical considerations. To his knowledge, the author does not collaborate with authoritative states. We have neither attempted to pay a bribe to an official. Even if our work is theoretical, the attack described is serious and could affect many data subjects. 

%
%
%

\section{Countermeasures}\label{sec:counter}

The verification made by the DPO to verify a request need to be relevant to the controller's knowledge of the data subject. Actually, document-based authentication can establish the identity of the subject securely even against governmental adversaries if three conditions are met: 
\begin{enumerate}
\item The data controller knows the identity of the subject and an identifying element (picture of the face or another biometrics).
\item The data controller and the subject sending the request are met physically. 
\item The DPO is trained to verify official documents (passport or ID card). 
\end{enumerate} 
If often fails because at least one of these conditions is not satisfied. If condition 3 is verified it can easily. The DPO needs to be trained to detect fake identification documents and document tampering. If the other conditions (1 and 2) are not verified, the patch is more costly.

\paragraph{Remote identity Proofing --} Let us assume that condition~2 cannot be satisfied, \textit{i.e.} the DPO and the subject cannot met physically like in a police control. The DPO needs to use a remote identity proofing scheme as described in~\cite{ENISA2021,ENISA2022}. The verification can be done using a video call or pictures. It relies on a liveness check and authenticity check (see~\cite{ENISA2021} for more details). it is important that the data controller knows both the identity and one biometrics of the subject before the verification. Otherwise a governmental adversary can succeed a forgery attack. If Bob knows only the identity of Alice, Charlie can forge a passport in the name of Alice with the picture and the biometrics of Eve. Eve can then impersonate Alice to Bob. Another approach for remote identity proofing is to rely on secure public electronic identification~\cite{ENISA2009}. Such systems are considered for deployment in the European Union. They could be used straightforward to authenticate subject access requests sent by European residents. However, data controllers would need to find another method to authenticate requests from data subjects which are outside the European Union.

\paragraph{Multi-factor Authentication --} MFA~\cite{Tilborg2011} is a solution to use if condition~1 cannot be satisfied. The DPO needs to consider all the information known on the subject to authenticate the subject's request. It can be a confirmation code sent on an email address or an SMS sent to a phone number for instance. It is important to notice that using a factor unknown by the controller before the request is not relevant and it cannot be trusted. 

If the data controller needs strong authentication to interact with the subject, then the authentication mechanism can be re-used to verify subject access requests. Both data processing and exercise of the subjects right aligned. An important question rises if the data controller does not need strong authentication to process subjects data. Does the data controller need to implement strong authentication to verify subject access requests? If we consider the use of multi-factor authentication, a data controller can ask the subject to provide a phone number when he/she registers for the sole purpose of authenticating later subject access requests. Recital 64 of the GDPR states that  \textit{A controller should not retain personal data for the sole purpose of being able to react to potential requests.} The data controller needs to implement the data minimization principle during the verification. There is a conflict between the application of the minimization principle and the security of subject access requests. Asking additional data to the subject seems necessary to have a safe exercise of the right to access.

\section{Conclusion}\label{sec:conclusion}

Governments have opportunities to abuse the right to access when a data controller authenticates subject access requests using only official identification documents. This form of authentication  is particularly weak against forgery attacks. Dissidents who use online services proposed by European companies can be targeted by authoritative governments they oppose. The scope of the GDPR makes our attack reach more people than expected. Forgery attacks can be even obtained by corruption.  If such attacks occur, these attacks would not only harm data subjects but they would also harm the trust on the personal data regulations like the GDPR. 

Finding a solution is not simple because data controllers need to find the good trade-off between security and privacy. Implementing strong authentication using multiple factors mitigates the risks related to handling subject access requests. But the data controller may need to obtain more information like the subject phone number or an email address to implement multi-factor authentication. Such information may not need necessary for the controller's processing. It is hard to satisfy both the data minimization principle (Recital 64 of the GDPR) and a useful and secure right to access. Data protection authorities and board need to clarify if minimization or right to access prevails.

\bibliographystyle{splncs04}
\bibliography{risky}

\end{document}